\documentclass[conference,10pt,a4paper]{IEEEtran}

\ifCLASSINFOpdf
\else
\fi
\usepackage{geometry}
\usepackage{amsthm,amssymb}
\usepackage[cmex10]{amsmath}

\usepackage{tikz}
\usetikzlibrary{shapes,arrows}
\usetikzlibrary{decorations.pathreplacing}
\usetikzlibrary{matrix}
\graphicspath{{./categoryFigure/}}

\usepackage{algorithmicx}
\usepackage{booktabs}
\usepackage[caption=false,font=footnotesize]{subfig}
\usepackage{fixltx2e}
\usepackage{graphicx} 
\usepackage{subfig}

\usepackage{multirow,multicol}
\usepackage{times}
\usepackage{bbm}

\newcommand{\bX}{\boldsymbol{X}} 
\newcommand{\balpha}{\boldsymbol{\alpha}} 
\usepackage{gensymb}
\makeatletter
\IEEEtriggercmd{\reset@font\normalfont\fontsize{9pt}{11pt}\selectfont}
\makeatother
\IEEEtriggeratref{1}
\IEEEoverridecommandlockouts
\usepackage{tikz}
\usepackage{textcomp}
\usepackage{hyperref}
\usepackage{lipsum}
\begin{document}
	\title{\Large \bfseries Phase I analysis of hidden operating status for wind turbine}
	\DeclareRobustCommand*{\IEEEauthorrefmark}[1]{%
		\raisebox{0pt}[0pt][0pt]{\textsuperscript{\footnotesize #1}}%
	}
	\author{\large \IEEEauthorblockN{Yuchen SHI\thanks{This research is conducted under
				Future Resilient Systems programme at Singapore-ETH
				Centre, established by ETH Z$\ddot{\textnormal{u}}$rich and Singapore's National Research Foundation (FI 370074011).}, Nan CHEN }
		\IEEEauthorblockA{Department of Industrial and Systems Engineering, National
			University of Singapore, Singapore\\(yuchen.shi@u.nus.edu, isecn@nus.edu.sg)} 
	}

\onecolumn{
	\Huge{IEEE Copyright Notice}\\
	\\
	\\
	\large{\textcopyright 2019 IEEE. Personal use of this material is permitted.
Permission from IEEE must be obtained for all other uses, in any current or future
media, including reprinting/republishing this material for advertising or promotional
purposes, creating new collective works, for resale or redistribution to servers or
lists, or reuse of any copyrighted component of this work in other works.}
\\
\\
\\
\\
\\
\\
\\
\large{Accepted to be published in: IEEE International Conference on Industrial Engineering and Engineering Management (IEEM 2019), Dec 15-18, 2019, Macao, China}}
\twocolumn
	\maketitle
    \begin{abstract}
	\bfseries Data-driven methods based on Supervisory Control and Data Acquisition (SCADA) becomes a recent trend for wind turbine condition monitoring. However, SCADA data are known to be of low quality due to low sampling frequency and complex turbine working dynamics. In this work, we focus on the phase I analysis of SCADA data to better understand turbines' operating status. As one of the most important characterization, the power curve is used as a benchmark to represent normal performance. A powerful distribution-free control chart is applied after the power generation is adjusted by an accurate power curve model, which explicitly takes into account the known factors that can affect turbines' performance. Informative out-of-control segments have been revealed in real field case studies. This phase I analysis can help improve wind turbine's monitoring, reliability, and maintenance for a smarter wind energy system. 
\end{abstract}

\medskip

\section{Introduction}
\label{sec:introduction}
Wind power is a competitive renewable energy source with increasing global capacity. However, current marginal profits of the wind industry are thin due to high operation and maintenance (OM) cost. Condition monitoring (CM) is one of the essential methodologies to help improve OM. With the fast development in computing power and data management capability, a recent trend of wind turbine (WT) condition monitoring is to interpret data from the Supervisory Control and Data Acquisition (SCADA) systems. A SCADA system supervises the wind farm by connecting the individual WTs, the substation and the meteorological stations to a central computer. It keeps a record of a few informative parameters on a 10-minute basis and implements control requirements for wind farm operations. Data-driven condition monitoring using SCADA data has the advantages of being cost-effective (no need for extra sensors), comprehensive (integrating subsystems of different mechanisms) and universally suitable for various hardware configurations (using data-driven methods). On the other hand, one of the most important characterizations to monitor WTs' dynamic performance is the \textit{power curve}. Generally speaking, the power curve depicts the relationship between power output and weather conditions. Related weather conditions include (but not limited to): wind speed, wind direction, air density, temperature, humidity, turbulence intensity. The impact of different weather conditions on power output is complicated, and thus, the power curve has no closed form. Site-specific power curves are recommended.  Based on the above discussions, in this paper, we consider the power curve estimated from SCADA data as a reference to construct condition monitoring of WTs. CM is usually implemented in two phases. To be more specific, this work focus on the Phase I analysis, which analyzes historical data retrospectively to characterize the in-control (IC) state and filter out variabilities in the data. Accurate IC status separation is an essential prerequisite for successful online monitoring model of incoming data (Phase II).

The real data from SCADA is known to contain failures (e.g., pitch malfunction, dirt/icing on blades), control adjustments (e.g., down rating), and sensor reading/ data transmission errors. Some cases can roughly be separated by alarm/control records in the SCADA data when these data are available \cite{park2014development,kusiak2011prediction}. When relevant information is not available, rough filtering can be done by clustering or robust data analysis techniques \cite{sainz2009robust,yampikulsakul2014condition}. Nevertheless, counting on SCADA alarm/control records or filtering of large deviations is not adequate due to the following reasons. Firstly, current warnings/alarms usually requires consecutive exceeding of limits. These requirements may miss the initial sign of faults and notable transit state on the turbine's performance \cite{park2014development}. Secondly, The real power curve is complex. It is not trivial for filtering techniques to accurately separate minimal deviations. It is even harder to filter out anomalies that do not show obvious deviations but only patterns ( see Fig. \ref{ab} as an example). Thirdly, the system status may change in the historical data, such as degradation. Awareness of degradation can help improve turbines' reliability and on-site power curve estimation \cite{sohoni2016critical}.  

Therefore, in this work, we aim to do a more thorough phase I analysis of SCADA data and better understand WT's operating status. To maximize its effectiveness, we recommend doing our phase I analysis after excluding the anomaly data that can be indicated by the alarm/control records or obvious deviations. Excluding these anomalies can effectively increase the robustness of the algorithm and reduce the masking effect.

There are several challenges of phase I analysis of SCADA data. Firstly, the power curve requires accurate estimation. \cite{lydia2014comprehensive} did a comprehensive review of the early efforts on power curve estimation. Recently, the superiority of nonparametric methods has been addressed by \cite{ding2015data}. We adopt one appealing nonparametric method in this work. Secondly, the residuals after power curve adjustments follow an unknown distribution, and possible anomalies include a board class of out-of-control (OC) patterns. For example, degradation can be considered as a single step change. A segment of down rating can be considered as a multi-steps change. Measurement errors can be considered as isolated changes. There are several nonparametric/distribution free control charts for phase I analysis in the literature \cite{chakraborti2001nonparametric,jones2009distribution,graham2010phase}. Control chart that is powerful at a board class of OC patterns is more appealing in this case. Thirdly, the residuals after power curve adjustments show positive serial correlations. Ignoring the serial correlation can lead to less efficient power curve estimation as well as less effective control charts. More importantly, to the extent of our knowledge, there is no control chart with effective performance under serial correlation. Therefore, reduction of autocorrelation is a crucial part of a successful phase I analysis. 

The method we propose in this work contains mainly two parts: estimating power curve from the SCADA data and checking the stability of the adjusted power generation process using a control chart. For the power curve estimation part, we adopt Multivariate Adaptive Regression Splines (MARS) \cite{friedman1991multivariate}. It is a spline-based flexible nonparametric regression technique that can deal with high dimensional interactive input, which makes it competitive when estimating power curve. MARS can explicitly takes into account the available variables that can affect power generation. Although many other nonparametric and machine learning methods, such as random forest and boosting, are able to deal with high dimensional data, a key appealing property of MARS is that the coefficients of spline basis functions are estimated by least square approach. Iterative Feasible Generalized Least Squares (IFGLS) can be naturally used to reduce the autocorrelation of the error term. In the end, a distribution-free control chart is conducted on the independent power curve residual. We adopt Recursive Segmentation and Permutation (RS/P) from \cite{capizzi2013phase}. RS/P has two major advantages. Firstly, it requires no knowledge about the IC distribution to achieve a relatively accurate preset false-alarm rate. Secondly, it is able to detect a board class of OC patterns such as isolations, single step changes, multi-step changes, and linear trend changes.

The remainder of this article is organized as follows. Section II explains the proposed phase I analysis methodology. In section III, we discuss the case study results of the proposed method on an open wind farm dataset. Section IV summarizes the article with potential future directions.

\section{Methodology}
\label{sec:method}
\subsection{MARS with IFGLS for power curve estimation}
\label{sec:mars}
Suppose we have dataset $\{(\bX_t,Y_{t}),t=1,\cdots,T\}$. Assume the following model for WT power curve as
\begin{equation} \label{model}
Y_{t}=f(\bX_t)+u_t,
\end{equation}
where $f(\cdot)$ captures the joint predictive relationship of $Y_{t}$ on $\bX_t$. $u_t$ is an autocorrelated but stationary process with mean 0. $u_{t}$ can be expressed in the autoregressive representation with order $p$
\begin{equation} \label{model_r}
u_{t}=\sum_{\theta=1}^{p} a_\theta u_{t-\theta}+\epsilon_{t},
\end{equation}
where $\epsilon_t$ are independently and identically distributed (i.i.d) with mean zero. $\epsilon_{t}$ follow unknown distribution. When the error term is independent, the smooth function $f(\bX_i)$ can be estimated by MARS using a set of basis functions
\begin{equation}
\hat{f}(\bX_t)=\sum_{s=1}^{S} \alpha_{s} B_{s}(\bX_t),
\end{equation}
where the basis functions take the form of
\begin{equation}
B_{s}(\bX_t)=\prod_{q=1}^{Q_{s}}\left[\eta_{q s} \cdot\left(X_{t,v(q, s)}-\gamma_{q s}\right)\right]_{+}.
\end{equation}
Here $Q_s$ is the number of knots. $v(q,s)$ denotes the input variable used by the $q$th multiplier in the $s$th basis function. $\eta_{qm}$ is a parameter with possible value $\pm1$. $\gamma_{q m}$ denotes the correspond knot location. All parameters $\{M,\alpha_s, Q_s, \eta_{qs},\gamma_{qs}, v_{(q,s)}\}$ are automatically determined by the data. The training procedure derives from recursive partitioning with extra efforts to ensure continuity. The lack-of-fit criterion used is the generalized cross-validation (GCV)
\begin{equation}\label{gcv}
\operatorname{GCV}(S)=\frac{1}{T} \sum_{i=1}^{T}\left[Y_{t}-\hat{f}\left(\bX_{t}\right)\right]^{2} /\left[1-\frac{C(S)}{N}\right]^{2},
\end{equation}
where 
\begin{equation*}
C(S)=\operatorname{trace}\left(\boldsymbol{B}\left(\boldsymbol{B}^{T} \boldsymbol{B}\right)^{-1} \boldsymbol{B}^{T}\right)+d\times S+1.
\end{equation*}
Here $\boldsymbol{B}$ is the $S\times T$ data matrix of the $S$ basis functions. $d$ represents a cost for each basis function with recommended value $d=2$.

In the end, the chosen basis functions are denote as $\boldsymbol{B}_{S^*}(\bX_t)=[B_{1}(\bX_t),\cdots,B_{S^*}(\bX_t)]^T$ with estimated coefficients $\hat{\balpha}_{S^*}=[\hat{\alpha}_1,\hat{\alpha}_2,\cdots,\hat{\alpha}_{S^*}]^T$. Similarly, denote the $S^*\times T$ data matrix of the $S^*$ basis functions as $\boldsymbol{B}_S^*$. 

Presence of autocorrelated observations makes it hard for phase I control charts to separate IC and OC points effectively. Therefore, we reduce autocorrelation before applying the control chart. By \eqref{gcv}, after $\boldsymbol{B}_{S^*}(\bX_t)$ is chosen, MARS estimate $\hat{\balpha}_{S^*}$ by the least square approach, which provides the attractive computational properties for us to reduce the autocorrelation in error term $u_t$. To be more specific, we can follow the IFGLS method, which is also known as Cochrane-Orcutt regression, using the following two steps. It is worth noticing that during this iteration, we always fix the spline basis functions as the one chosen by MARS ($\boldsymbol{B}_{S^*}(\bX_t)$). Only $\hat{\balpha}_{S^*}$ are updated.

\noindent\textbf{Step 1: Estimate AR parameters given $\hat f(\cdot)$}

Given existing estimates $\hat{\balpha}_{S^*}$ and the mean function $\hat f(\bX_t)=\boldsymbol{B}_{S^*}(\bX_t)^T\hat{\balpha}_{S^*}$ for every $t$, estimate the parameters $\{{a}_1,\cdots {a}_p \}$ associated with the autoregressive
process $\hat{u}_t=[Y_t-\hat{f}(\bX_t)]$.. With reasonable $\hat
f(\bX_t)$, $\hat u_t$ is a stationary process and thus conventional time series
techniques can be used. To account for flexible distribution of $\epsilon_t$, we
choose the least square approach.
\begin{equation*}
\hat{a}_1,\cdots\hat{a}_p=\arg\min_{{a}_1,\cdots {a}_p}
\sum_{t=p+1}^T\left[\hat u_t-a_1\hat{u}_{t-1}-\cdots- {a}_{p}\hat{u}_{t-p}\right]^2
\end{equation*}
The order $p$ is a tuning parameter, and can be adaptively determined by model selection criteria.
\\\noindent\textbf{Step 2: Update $\hat{\balpha}_{S^*}$ given AR parameters}

Substitute $\hat{Y}_t=Y_t-\sum_{\theta=1}^{p}\hat{a}_\theta\hat u_{t-\theta}$ and update $\hat{\balpha}_{S^*}$ using $\hat{Y}_t$ by least square approach,
\begin{equation*}
\hat{\balpha}_{S^*}=(\boldsymbol{B}_{S^*}\boldsymbol{B}_{S^*}^T)^{-1}\boldsymbol{B}_{S^*}\hat{\boldsymbol{Y}},
\end{equation*}
where $\hat{\boldsymbol{Y}}=[\hat{Y}_1,\hat{Y}_2,\cdots,\hat{Y}_T]^T$.
In this way, we can have
\begin{equation}
\mathbb{E}[Y_t|\boldsymbol{X}_t]=\boldsymbol{B}_{S^*}^T(\boldsymbol{X}_t)\hat{\balpha}_{S^*}+\sum_{\theta=1}^{p} \hat{a}_\theta u_{t-\theta},
\end{equation}
\begin{equation}
r_t=Y_t-\mathbb{E}[Y_t|\boldsymbol{X}_t]
\end{equation}
We set the converge threshold of $\hat{a}_1,\cdots\hat{a}_p$ to be 0.001. At the same time, we do Box-Ljung test on all the $p$ lags of the series $r_t$. The iteration terminates when the AR parameters converge and all $p$ lags pass the Box-Ljung test, which indicate that $r_t$ can be treated as independent.

\subsection{RS/P for phase I analysis of SCADA data}
\label{sec: rsp}
If the process is in control, $r_t$ can be treated as i.i.d. Therefore, control statistics can be constructed based on $r_t$. Due to the complexity of WTs' operation, $r_t$ does not follow a normal distribution. Besides, it is risky to assume any known distribution form before the process is stable. Therefore, we adopt the powerful distribution-free phase I analysis (RS/P) proposed by \cite{capizzi2013phase}. As explained in section I, RS/P offers a satisfactory performance against a board class of out-of-control (OC) patterns without any knowledge about the IC distribution. In this work, we only focus on level-changes detection, i.e., the changes in mean values of the $r_t$ process. This is because wind itself is a stochastic process. Scale changes of $r_t$ severely depend on the patterns of arriving wind. Scale-changes detection is non-trivial and requires further developments.

Group $r_t$ into $m$ subgroups of size $n$ successive individual observations. Assume IC observations follow an unknown but common mean $\mu_0$. When the process is OC, assume the following multiple change-point model, 
\begin{equation}\label{OC}
r_{i j} \sim \left\{\begin{array}{ll}{\mu_{0}} & {\text { if } 0<i \leq \tau_{1}} \\ {\mu_{1}} & {\text { if } \tau_{1}<i \leq \tau_{2}} \\ {\vdots} & {\vdots} \\ {\mu_{k}} & {\text { if } \tau_{k}<i \leq m}\end{array}\right.
\end{equation}
where $i=1,\cdots,m$ and $j=1,\cdots,n$. $0<\tau_{1}<\tau_{2}<\cdots<\tau_{k}<m$ denote $k$ change points and $\mu_0 \cdots \mu_k$ are the mean values, which are all assume to be unknown. This OC model has the potential to cover a board range of OC patterns such as single step, multi steps and isolated changes. 

Based on \eqref{OC}, control statistics are computed separately for isolated shifts and $k=1,\cdots K$ step shifts. $K+1$ control statistics are then aggregated as an overall control statistic, which is used to compute the significance level.

The control statistic for isolated shifts is the standard Shewhart $\bar{X}$, $T_{0}=\max _{i=1, \ldots, m}\left|\overline{r}_{i}-\overline{\overline{r}}\right|$. Here $\overline{r}_{i}=\frac{1}{n} \sum_{j=1}^{n} r_{i j}$ and $ \overline{\overline{r}}=\frac{1}{m} \sum_{i=1}^{m} \overline{r}_{i}$.

The control statistics of step shifts are computed by recursive segmentation whereas the change points are located simultaneously. For step changes, we further assume the length between step changes are larger than $l_{\min}$. $l_{\min}$ is recommended to be 5 to reduce the IC variability of control statistics whereas remaining an effective detecting power. There are totally $K+1$ successive stages ($k=0,\cdots ,K$). At the beginning of each $k$, the interval $[1,m]$ has already been split into $k$ subintervals by the previous $k$ stages. At stage $k$, a new change point is chosen by 
\begin{equation*}\label{seg}
\hat{\tau}_{i}=\arg\max\sum_{i=1}^{k+1}\left(\hat{\tau}_{i}-\hat{\tau}_{i-1}\right)\left(\overline{r}\left(\hat{\tau}_{i-1}, \hat{\tau}_{i}\right)-\overline{\overline{r}}\right)^{2},
\end{equation*}
Where $\bar{r}(a, b)$ is the average of subgroup means between $a$ and $b$. Given $\hat{\tau}_{1},\cdots,\hat{\tau}_{k+1}$, $T_k=\max\sum_{i=1}^{k+1}\left(\hat{\tau}_{i}-\hat{\tau}_{i-1}\right)\left(\overline{r}\left(\hat{\tau}_{i-1}, \hat{\tau}_{i}\right)-\overline{\overline{r}}\right)^{2}$.

The significance level ($p$-value) of the control statistics is based on permutation. If $\boldsymbol{r}= \{r_1,r_2,\cdots, r_T\}$ is IC, we have
\begin{align*}
&P\left\{\boldsymbol{r}=\left(z_{1}, \ldots, z_{T}\right) | \boldsymbol{r}_{(\cdot)}\right\}    \\&=\left\{\begin{array}{ll}{\frac{1}{N !}} & {\text { if }\left(z_{1}, \ldots, z_{T}\right) \text { is a permutation of } \boldsymbol{r}_{(\cdot)}}, \\ {0} & {\text { otherwise. }}\end{array}\right.
\end{align*}
where $\boldsymbol{r}_{(\cdot)} $ is the order statistic of pooled sample of size $T$. Since $P\left\{\boldsymbol{r}=\left(z_{1}, \ldots, z_{T}\right) | \boldsymbol{r}_{(\cdot)}\right\}$ does not depend on the distribution of $r_t$, the $p$-value can be computed using permutations. To be more specific, we can generate $L$ random permutations of $\boldsymbol{r}$ and calculated $\tilde{T}_{kl}$ for every $l$ and $k$. The aggregated control statistics can be constructed by:
\begin{equation}\label{overall}
W=\max _{k=0, \ldots, K} \frac{T_{k}-u_{k}}{v_{k}},
\end{equation}
where $u_{k}=1/L\sum_{l=1}^{L}\widetilde{T}_{k l}$ and $v_{k}^{2}=1/(L-1) \sum_{l=1}^{L}\left(\widetilde{T}_{k l}-u_{k}\right)^{2}$. The $p$-value of $W$ can be determined as
\begin{equation}\label{p_control}
p=\frac{1}{L} \sum_{l=1}^{L} I\left(\widetilde{W}_{l} \geq W\right),
\end{equation}
where $\widetilde{W}_{l}=\max _{k=0, \ldots, K} (\widetilde{T}_{k l}-u_{k})/{v_{k}}$. $I$ denotes an indicator function. 

\begin{figure}[]
	\centering
	\includegraphics[width=0.96\linewidth]{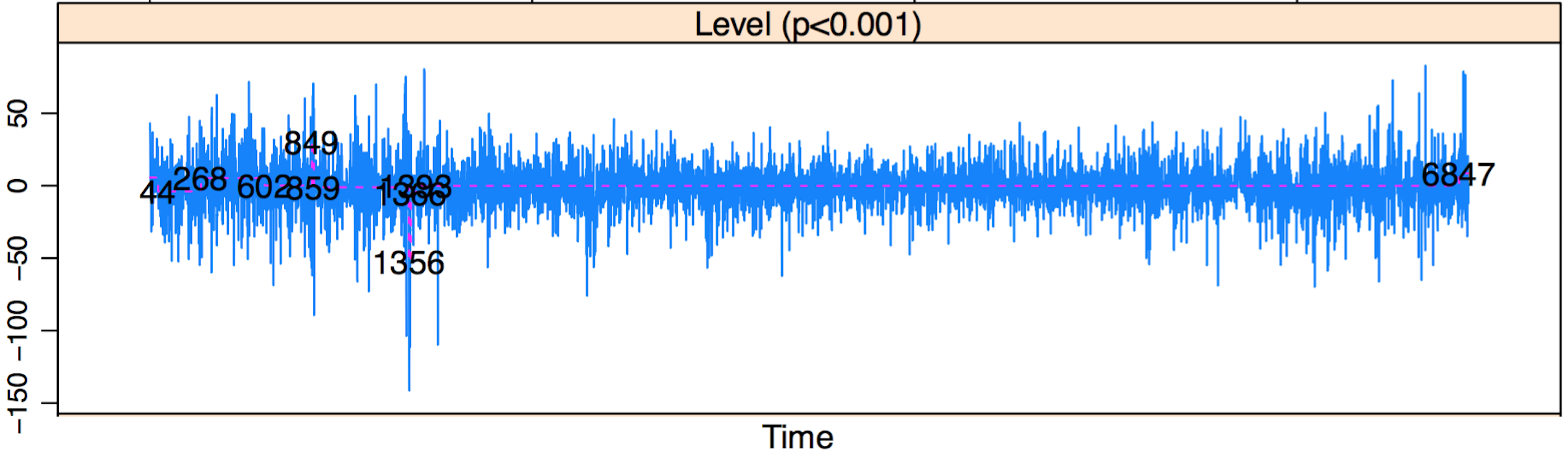}
	\caption{An example of phase I analysis using RS/P. The numbers marked on the chart denotes the change points $\tau_i$}\label{fig:diagnosis}
\end{figure}

Finally, since the model \eqref{OC} detects $k$ changes simultaneously, as is shown in Fig. \ref{fig:diagnosis}, in the real application, we remove one detected OC segments with the largest shifted mean every time and re-do the phase I analysis. The iteration terminates when the significance level $p$ in \eqref{p_control} increase to be larger than a preset threshold. 
\section{Experiments}
\label{sec:experiment}
The case study datasets are from an open data wind farm, ``La Haute Borne" (Meuse, France), provided by ENGIE Renewable Energy\footnote{The dataset used in this case study is available at https://opendata-renewables.engie.com/pages/home/}. The dataset contains 10-min average historical data of 4 WTs from year 2013 to now and we show experiments on turbine ``R80711". Available variables include weather conditions and turbine's responses yet no control/alarm records. Therefore, before phase I analysis, we first do rough filtering to separate the variabilities that can be indicated by the recorded variables using the following rules:
\begin{itemize}
	\item Remove the points with power output smaller or equal to 0 kW, which are considered to be at ``idle" state. 
	\item Remove points that are right next to ``idle" state, which are at ``startup/shutting down" state. 
	\item Remove points with pitch angle that is higher than 20$^{\circ}$, which indicate pitch control. 
\end{itemize}
We conduct two experiments with different data length. Information about the two datasets is shown in Table \ref{exper}.

\begin{figure}[t]
	\centering
	\subfloat[original data]{\includegraphics[width=0.48\linewidth]{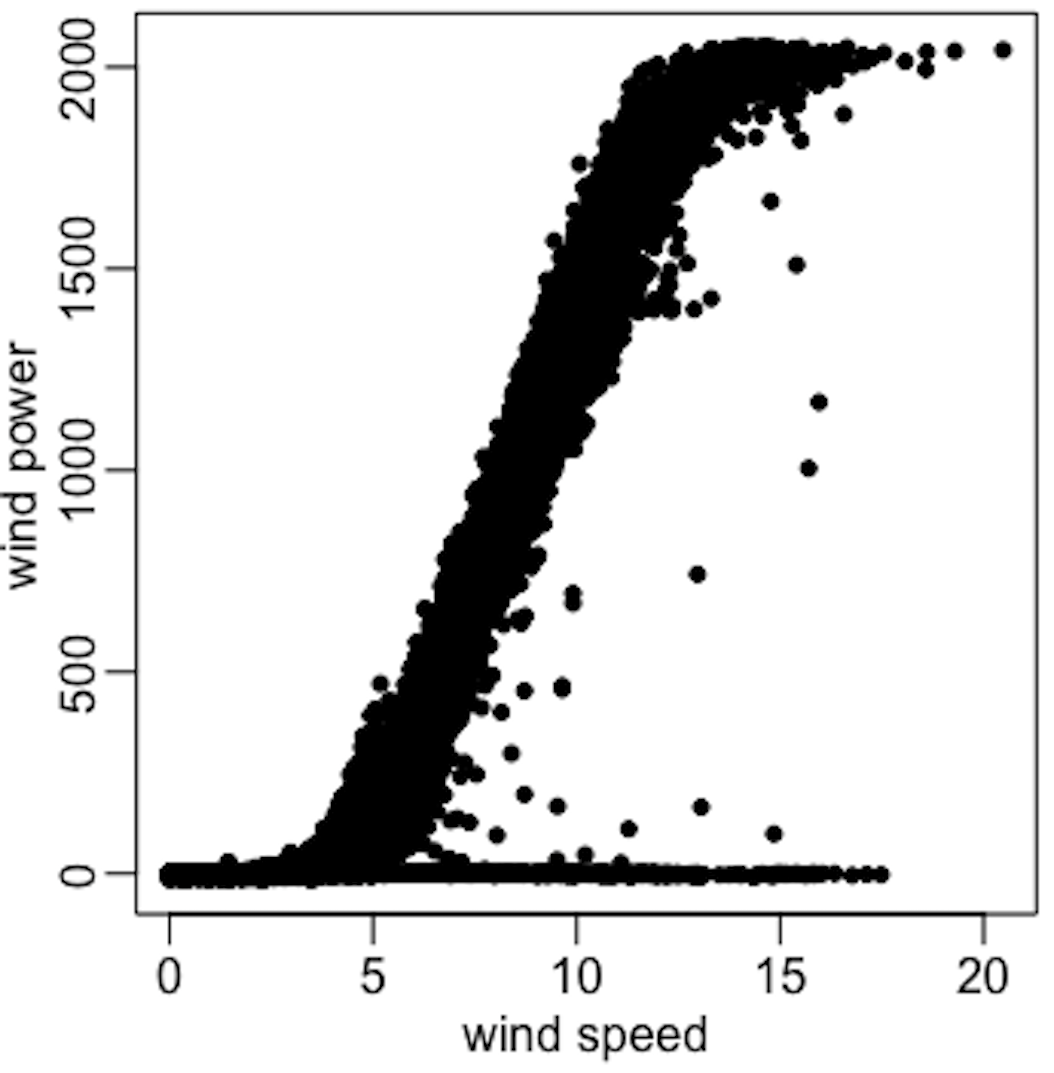}}
	\subfloat[data after rough filtering]{\includegraphics[width=0.48\linewidth]{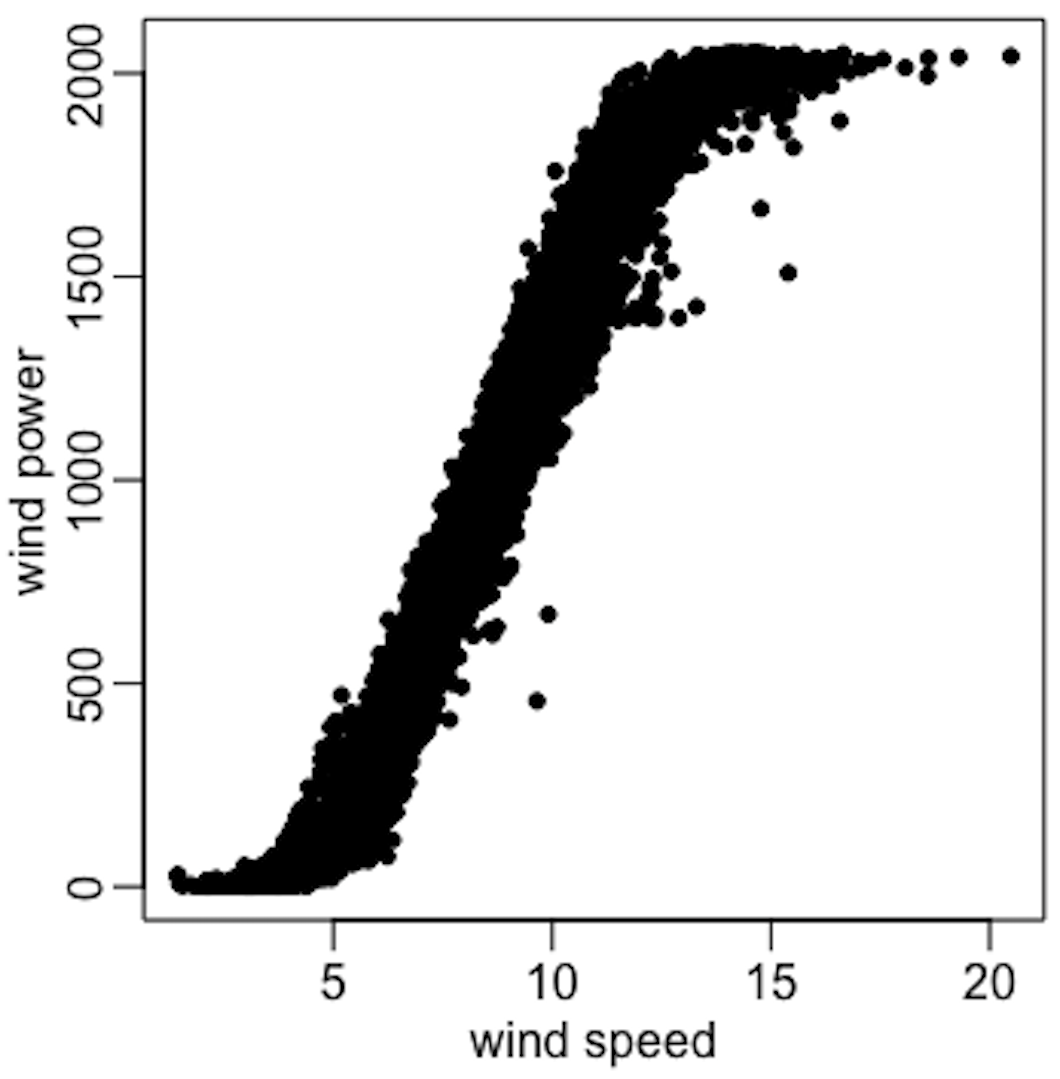}}
	\caption{Rough filtering of La Haute Borne SCADA data. Left panel: original data; Right panel: data after rough filtering}\label{fig:original}    
\end{figure}

We plot the original dataset 2 and dataset 2 after rough filtering as an example. It can be seen from Fig. \ref{fig:original} that after rough filtering, there is no obvious deviation from normal power curve. Our proposed phase I analysis can therefore be conducted. To apply the proposed model in section II, we implement the 'mars' function in R package 'mda' for MARS. The RS/P control chart is implemented by the 'rsp' function in package 'dfphase1'. 

Based on the variables available, when estimating MARS, we use the 10-min average records of all the available weather condition variables: wind speed, wind direction, outdoor temperature, and turbulence intensity. Since 10-min is a coarse interval, we also include the standard deviation of wind direction and outdoor temperature to reduce information loss. The standard deviation of wind speed is not added since it is covered by turbulence intensity. Last but not least, we also include ``month" to account for the seasonal effect. Root Mean Squared Error (RMSE) is chosen as the measure of accuracy, the result of MARS and MARS with Iterative FGLS are shown in Table \ref{fitting}, which shows the effectiveness of reducing autocorrelation.

\begin{table}[t]
	\centering
	\caption{Experiments}\label{exper}    
	\begin{tabular}{p{0.05\linewidth}p{0.54\linewidth}p{0.25\linewidth}}
		\toprule        
		&Datasets&Number of Points after filtering\\
		\hline
		1&2012-12-31 23:00 to 2013-04-01 22:00& 10014\\  
		2&2016-01-01 23:00 to 2016-12-31 22:50&41374 \\       
		\bottomrule
	\end{tabular}
\end{table}
\begin{table}[t]
	\centering
	\caption{RMSE of the fitted power curve}\label{fitting}    
	\begin{tabular}{p{0.1\linewidth}p{0.15\linewidth}p{0.35\linewidth}}
		\toprule        
		Dataset &MARS&MARS with IFGLS\\
		\hline
		1& 39.18 & 30.08\\
		2& 42.86 & 35.84\\
		\bottomrule
	\end{tabular}
\end{table}
In our case study, we subgroup the points hourly with $n=6$. Set maximum number of change points detected $k=50$ and significance level $p=0.05$. Besides, since dataset 2 has a large sample size, we only plot the first 4 segments detected. Excluding these 4 segments is able to increase the significance level of aggregated control statistic in equation \eqref{p_control} 0.002. Excluding another 4 more segments is able to increase the $p-$ value to be larger than 0.05.
\begin{figure}
	\centering
	\subfloat[OC segments of dataset1]{\includegraphics[width=0.96\linewidth]{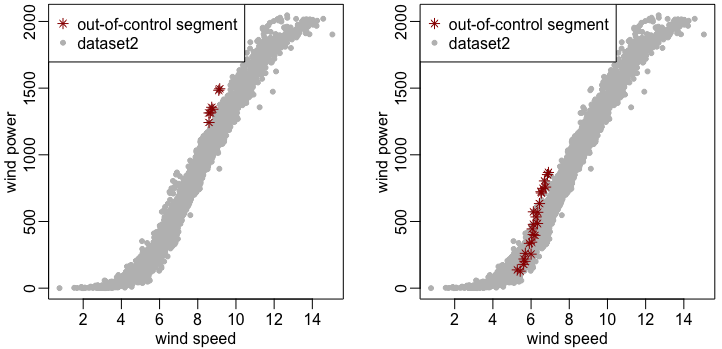}}\\
	\subfloat[OC segments 1-2 of dataset 2]{\includegraphics[width=0.96\linewidth]{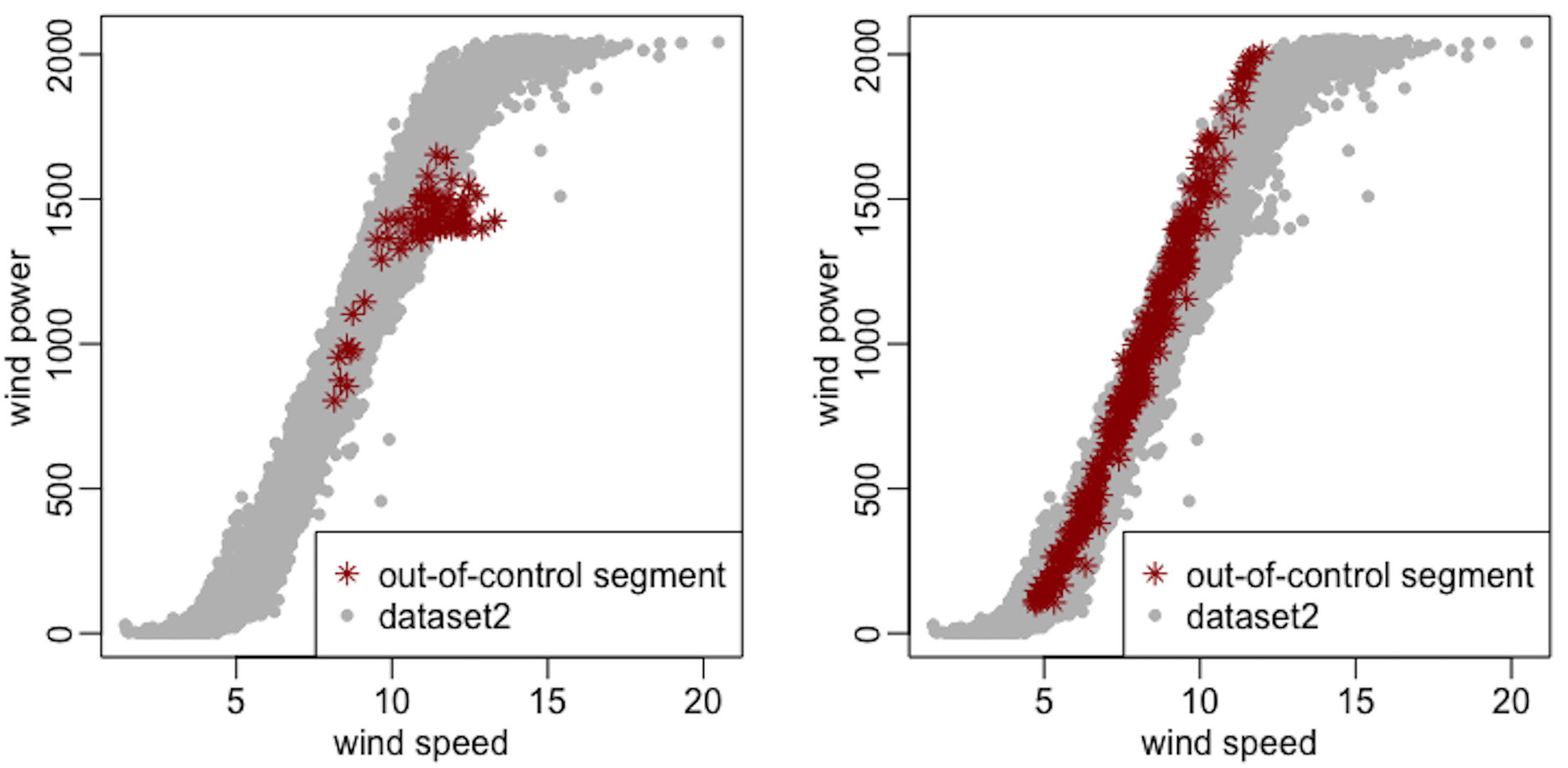}}\\
	\subfloat[OC segments 3-4 of dataset 2]{\includegraphics[width=0.96\linewidth]{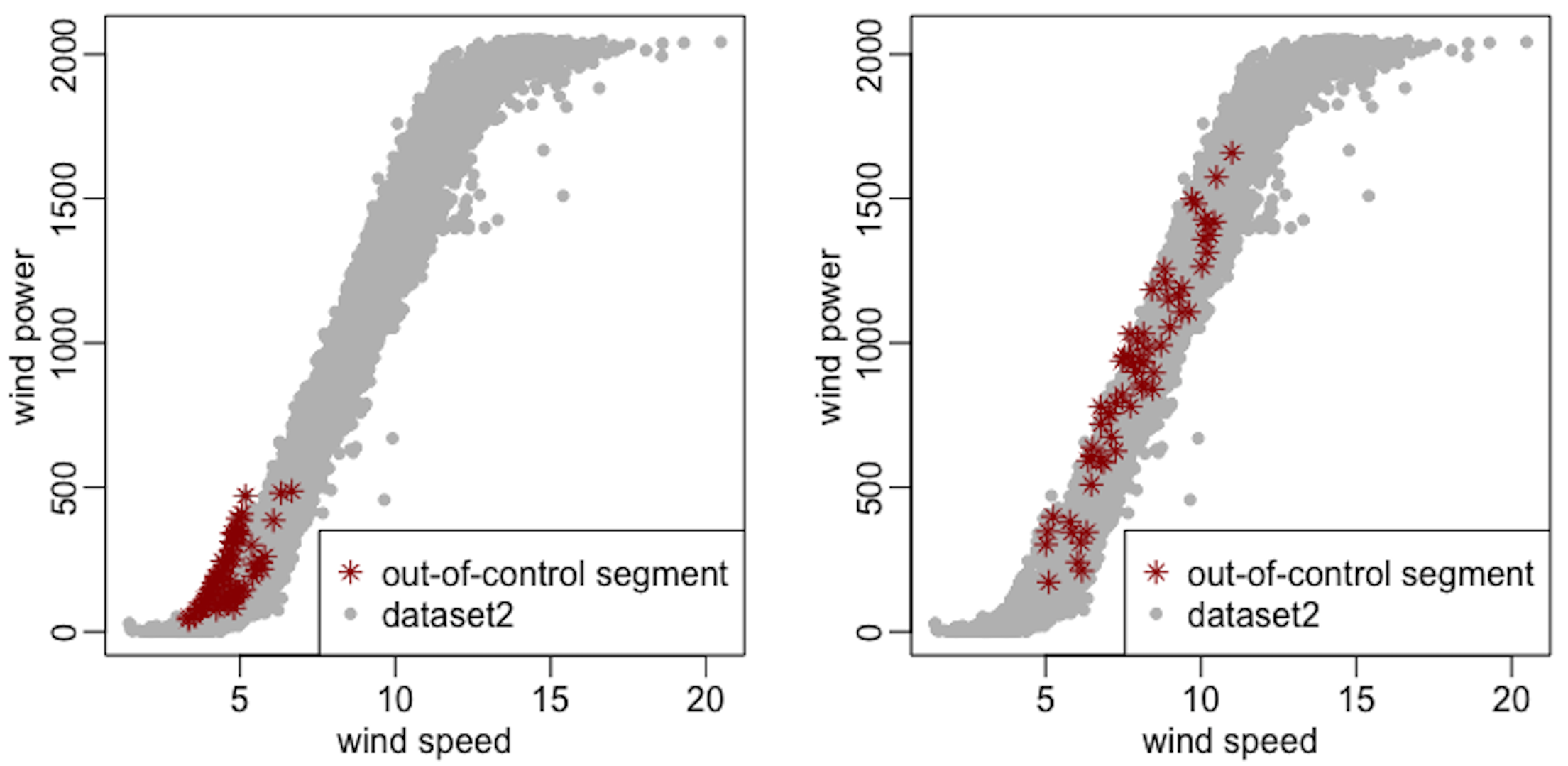}}\\    
	\caption{Detection OC segments based on power curve}\label{ab}
\end{figure}

Fig. \ref{ab} plots the detected OC data. The detected OC segments show variations of pattern compared to normal data. To be more specific, the left panel of (a) and left panel of (c) shows a short period of deviation larger than the power curve. A possible situation is bad sensor reading or inefficient SCADA data transmission. Right panel if (a) and right panel of (b) shows a higher mean function slope. A possible situation is increasing of air density. Air density affects power generation yet is not available in this dataset. Left panel of (b) shows a segment concentrated at mean values much lower than expected. This is probably a down-rating period due to power integration adjustments or wake effects alleviation. Last but not least, Right panel of (c) shows a larger variance. This segment is from 2016-02-09 15:20 to 2016-02-10 10:20, among which points among 2016-02-10 00:40 to 09:20 are filtered out by rough filtering because of high pitch angle value (stabilized at 90$^{\circ}$). Such a high pitch angle at lower speed range is usually set for emergency shut down. The segment in the right panel of (c) is mostly the segment right before an emergency shut down. It could be a segment under sub-optimal operation.  

Last but not least, it is non-trivial to show the effectiveness of phase I control chart in case studies, since the precise system status can not be verified given available records. Nevertheless, \cite{capizzi2013phase} has shown by simulation that RS/P performs globally better than the competitive nonparametric phase I analysis in the literature under various OC scenarios \cite{jones2009distribution}, \cite{graham2010phase}. To be more specific, RS/P always performs better for patterned shift and performs at least comparable for isolated shifts. RS/P also performs comparable regarding to IC performance. Moreover, the OC segments detected above are unable to be detected by clustering or robust regressions, since they show patterns instead of obvious deviations. At the same time, the autocorrelation and unknown distribution of power curve mean residual ensure the effectiveness of our phase I analysis. Ignoring autocorrelation and wrong distributional assumption both increase the false alarm rate. 
 
\section{Conclusion}
 In this work, we propose a phase I analysis of SCADA data to better understand WT's operating status. We firstly build a power curve model that explicitly accounts for the available factors to represent WT's normal performance. A distribution-free control chart (RS/P) is then adopted to detect level changes on the error term. The case studies show a variety of informative OC patterns detected. Given the currently available information, we provide possible explanations for the OC patterns as insights. Nevertheless, more considerate diagnoses should be further conducted with more information and expert knowledge. This phase I analysis shows the potential to better understand wind turbine's operating status and thus improves wind turbines' monitoring, reliability, and maintenance for a smarter wind energy system. For further studies, more work needs to be done on detecting scale shifts, which also contains valuable information. Also, in this work, we did not monitor the coefficients in the power curve model. Both are non-trivial and require further developments. 
\label{sec:conclusion} 
\bibliographystyle{IEEEtran}
\bibliography{phase1_literature}

\begin{thebibliography}{10}
\providecommand{\url}[1]{#1}
\csname url@samestyle\endcsname
\providecommand{\newblock}{\relax}
\providecommand{\bibinfo}[2]{#2}
\providecommand{\BIBentrySTDinterwordspacing}{\spaceskip=0pt\relax}
\providecommand{\BIBentryALTinterwordstretchfactor}{4}
\providecommand{\BIBentryALTinterwordspacing}{\spaceskip=\fontdimen2\font plus
\BIBentryALTinterwordstretchfactor\fontdimen3\font minus
  \fontdimen4\font\relax}
\providecommand{\BIBforeignlanguage}[2]{{%
\expandafter\ifx\csname l@#1\endcsname\relax
\typeout{** WARNING: IEEEtran.bst: No hyphenation pattern has been}%
\typeout{** loaded for the language `#1'. Using the pattern for}%
\typeout{** the default language instead.}%
\else
\language=\csname l@#1\endcsname
\fi
#2}}
\providecommand{\BIBdecl}{\relax}
\BIBdecl

\bibitem{park2014development}
J.-Y. Park, J.-K. Lee, K.-Y. Oh, and J.-S. Lee, ``Development of a novel power
  curve monitoring method for wind turbines and its field tests,'' \emph{IEEE
  Transactions on Energy Conversion}, vol.~29, no.~1, pp. 119--128, 2014.

\bibitem{kusiak2011prediction}
A.~Kusiak and A.~Verma, ``Prediction of status patterns of wind turbines: A
  data-mining approach,'' \emph{Journal of Solar Energy Engineering}, vol. 133,
  no.~1, p. 011008, 2011.

\bibitem{sainz2009robust}
E.~Sainz, A.~Llombart, and J.~Guerrero, ``Robust filtering for the
  characterization of wind turbines: Improving its operation and maintenance,''
  \emph{Energy Conversion and Management}, vol.~50, no.~9, pp. 2136--2147,
  2009.

\bibitem{yampikulsakul2014condition}
N.~Yampikulsakul, E.~Byon, S.~Huang, S.~Sheng, and M.~You, ``Condition
  monitoring of wind power system with nonparametric regression analysis,''
  \emph{IEEE Transactions on Energy Conversion}, vol.~29, no.~2, pp. 288--299,
  2014.

\bibitem{sohoni2016critical}
V.~Sohoni, S.~Gupta, and R.~Nema, ``A critical review on wind turbine power
  curve modelling techniques and their applications in wind based energy
  systems,'' \emph{Journal of Energy}, vol. 2016, 2016.

\bibitem{lydia2014comprehensive}
M.~Lydia, S.~S. Kumar, A.~I. Selvakumar, and G.~E.~P. Kumar, ``A comprehensive
  review on wind turbine power curve modeling techniques,'' \emph{Renewable and
  Sustainable Energy Reviews}, vol.~30, pp. 452--460, 2014.

\bibitem{ding2015data}
Y.~Ding, J.~Tang, and J.~Z. Huang, ``Data analytics methods for wind energy
  applications,'' in \emph{ASME Turbo Expo 2015: Turbine Technical Conference
  and Exposition}.\hskip 1em plus 0.5em minus 0.4em\relax American Society of
  Mechanical Engineers, 2015, pp. V009T46A020--V009T46A020.

\bibitem{chakraborti2001nonparametric}
S.~Chakraborti, P.~Van~der Laan, and S.~Bakir, ``Nonparametric control charts:
  an overview and some results,'' \emph{Journal of Quality Technology},
  vol.~33, no.~3, pp. 304--315, 2001.

\bibitem{jones2009distribution}
L.~A. Jones-Farmer, V.~Jordan, and C.~W. Champ, ``Distribution-free phase i
  control charts for subgroup location,'' \emph{Journal of Quality Technology},
  vol.~41, no.~3, pp. 304--316, 2009.

\bibitem{graham2010phase}
M.~Graham, S.~Human, and S.~Chakraborti, ``A phase i nonparametric
  shewhart-type control chart based on the median,'' \emph{Journal of Applied
  Statistics}, vol.~37, no.~11, pp. 1795--1813, 2010.

\bibitem{friedman1991multivariate}
J.~H. Friedman \emph{et~al.}, ``Multivariate adaptive regression splines,''
  \emph{The annals of statistics}, vol.~19, no.~1, pp. 1--67, 1991.

\bibitem{capizzi2013phase}
G.~Capizzi and G.~Masarotto, ``Phase i distribution-free analysis of univariate
  data,'' \emph{Journal of Quality Technology}, vol.~45, no.~3, pp. 273--284,
  2013.

\end{thebibliography}

\end{document}